\documentclass[prb,twocolumn,showpacs]{revtex4}
\usepackage{graphicx}
\usepackage{amssymb}
\usepackage{amsmath}
\usepackage{color}

\begin{document}

\title{Polarons and slow quantum phonons}

\author{Andreas Alvermann}
\author{Holger Fehske}
\affiliation{
Institut f{\"u}r Physik, Ernst-Moritz-Arndt-Universit{\"a}t Greifswald,
17489 Greifswald, Germany }
\author{Stuart A. Trugman}
\affiliation{Theoretical Division, Los Alamos National Laboratory, Los
Alamos, New Mexico 87545, USA}

\begin{abstract}

We describe the formation and properties of Holstein polarons in the entire parameter regime.
Our presentation focuses on the polaron mass and radius,
which we obtain with an improved numerical technique.
It is based on the combination of variational exact diagonalization with an improved construction of phonon states, providing results even for the strong coupling adiabatic regime.
In particular we can describe the formation of large and heavy adiabatic polarons.
A comparison of the polaron mass for the one and three dimensional situation 
explains how the different properties in the static oscillator limit determine the behavior in the adiabatic regime.
The transport properties of large and small polarons are characterized by the f-sum rule and the optical conductivity.
Our calculations are approximation-free and have negligible numerical error.
This allows us to give a conclusive and impartial description of polaron formation.
We finally discuss the implications of our results for situations beyond the Holstein model. 
\end{abstract}

\pacs{71.38.-k,71.38.Cn,71.38.Ht} 

\maketitle

\section{Introduction}

In many materials it is the strong coupling of electronic and lattice degrees of freedom that determines the charge transport or the optical properties~\cite{La33,Pe46b,Ra82,SS93,Alex07,AD09}.
A basic effect is the renormalization of the charge carrier mass~\cite{MG40}.
The mass renormalization is especially large in materials with a short-ranged interaction of electrons with optical phonons, as in molecular crystals or the colossal magnetoresistive manganites~\cite{HMLK06}.
In such materials it happens that charge carriers, which otherwise resemble free electrons apart from a certain mass renormalization,
give way to a kind of new quasiparticle with entirely different properties,
the Holstein polaron~\cite{Ho}.
The mass of Holstein polarons can exceed that of a Bloch electron by several orders of magnitude, and may thus explain a low conductivity or thermally activated transport~\cite{BB85}.
On the other hand, polarons have specific optical signatures~\cite{Fi95}.

From early work we understand the fundamental mechanism of polaron formation very well in two opposite cases:
In the antiadiabatic regime of large phonon frequencies, where strong coupling perturbation theory provides the picture of a Lang-Firsov polaron~\cite{LF}, and in the strict adiabatic limit of static oscillators, where the electron and phonon wavefunctions factor~\cite{EH,KM93}.

Much analytical and numerical work has been performed to extend our understanding to finite phonon frequencies and the intermediate regime,
including extended variational approaches~\cite{FILTB94,RBL98,Ba07}, exact diagonalization (ED)~\cite{St96,WF,MR97,BTB,BA02,DCMP05,LBBM10}, density matrix renormalization group~\cite{JW98b}, quantum Monte Carlo~\cite{RL,BVL95,Kor,MNAFDCS09},
and different perturbative or approximate schemes~\cite{Ea,CPFF97,RBL99,GBS06,BB08,Zo10}.
Numerical calculations provide reliable data for a wide range of phonon frequencies and coupling strengths, against which any discussion of Holstein polarons must be checked (for a recent review, see e.g.~Refs.~\onlinecite{FT07,JF07}).
It is however very difficult to perform comparable calculations
for small phonon frequencies, since the increasing number of phonon excitations
poses a severe restriction on any, even large-scale, numerics.
So far this prevented a thorough study of the 
experimentally most relevant adiabatic regime,
in particular of large and heavy polarons as predicted by adiabatic approximations for one-dimensional (1D) systems. Also, numerical data for optical or transport properties are scarce~\cite{SWWAF05,FC06}.

In the present contribution we use a new numerical technique to extend previous studies deep into the adiabatic regime.
The key idea is an improved construction of phonon states that allows for the simultaneous representation of small and large oscillator shifts with few elements in a Hilbert space. The construction overcomes the problem 
that the number of oscillator energy eigenstates needed grows fast at small frequency.
We combine this construction with the phonon state selection in variational exact diagonalization (VED)~\cite{BTB}.
Based on this method we present here results for phonon frequencies down to a hundredth of the electron transfer integral and still at large coupling,
extending the frequency range accessed in previous studies by more than one order of magnitude.
We explore the Holstein polaron in parameter regimes that, so far, were only accessible to approximate treatments.
The validity of common concepts, and thus our very picture of polaron formation, can now be assessed.
This includes the question which properties of the static limit remain at small but finite phonon frequencies, and how the limit itself is approached.
To our knowledge it is for the first time that the formation of large and heavy 1D polarons is studied with unbiased numerics.

The paper is organized as follows.
In Sec.~\ref{sec:model} we explain our numerical technique in application to the Holstein model.
The differences between the 1D and 3D case are discussed through the comparison of the polaron mass in Sec.~\ref{sec:mass}, and the polaron radius in Sec.~\ref{sec:radius}.
The study of the 1D case is continued with the f-sum rule and optical conductivity in Sec.~\ref{sec:1d}.
We conclude in Sec.~\ref{sec:conclusion} with a discussion of our results in the broader context of polaron formation under general conditions.

\section{The Holstein model and the numerical method}
\label{sec:model}

The Hamiltonian of the Holstein model on a chain (cubic lattice) in dimension $D=1$ ($D=3$) is given by
\begin{equation}\label{Ham}
\begin{split}
  H = & - t \sum_{i=1}^D \sum_\mathbf{n}  c^\dagger_{\mathbf{n}+
    \mathbf{e}_i} c^{}_\mathbf{n} + c^\dagger_\mathbf{n} c^{}_{\mathbf{n}+
    \mathbf{e}_i} \\
  & - \sqrt{\varepsilon_p\, \omega_0} \sum_\mathbf{n} (b^\dagger_\mathbf{n} +
  b^{}_\mathbf{n}) c^\dagger_\mathbf{n} c^{}_\mathbf{n} + \omega_0
  \sum_\mathbf{n} b^\dagger_\mathbf{n} b^{}_\mathbf{n} \;.
\end{split}
\end{equation}
The fermionic operators $c^\dagger_\mathbf{n}$ create the electron, and the bosonic operators $b^\dagger_\mathbf{n}$ create a phonon at site $\mathbf{n}$.
The sums run over all lattice sites $\mathbf{n}$, with the vectors $\mathbf{e}_i$ along the lattice axes.
The Holstein model has three parameters: the electron transfer integral (or hopping matrix element) $t$, the phonon frequency $\omega_0$, and a third parameter  $\epsilon_p$, $\lambda=\epsilon_p/2 D t$, or $g^2=\epsilon_p/\omega_0$ specifying the electron-phonon coupling strength.
In the following we give all numerical values with the hopping matrix element as the unit of energy, so that $t=1$.
All results are for a single electron at zero temperature.

Our calculations for the Holstein model are performed with standard ED methods, such as the Lanczos algorithm and the Kernel Polynomial Method~\cite{WWAF06}. The fundamental task is the construction of an appropriate subspace of the complete infinite Hilbert space of the single fermionic and many bosonic degrees of freedom of the Hamiltonian~\eqref{Ham}.
A very efficient construction is provided by VED,
which achieves extreme accuracy not only for the polaron groundstate, especially in the regime of intermediate coupling and phonon frequency~\cite{BTB}, but also for models with different types of fermion-boson coupling~\cite{AEF07}.

VED is based on an increasing sequence of subspaces of the complete Hilbert space.
A certain subspace contains all basis states that are produced through a given number $N_\mathrm{gen}$ of applications of the Hamiltonian to the phonon vacuum, that is either by excitation of a phonon or an electron move.
For example, it contains the state with $N_\mathrm{gen}$ phonons at the electron position, and the one with a single phonon  $N_\mathrm{gen}-1$ lattice sites away from the electron.  
The number of elements in the subspaces grow only by a factor close to $D+1$ as $N_\mathrm{gen}$ increases by $1$, and already small subspaces provide a good approximation of the true groundstate. All quantities, e.g. variational estimates of the groundstate energy, can be improved by increasing $N_\mathrm{gen}$ until convergence.

The main restriction of the VED construction, common to most ED studies,
is the large number of phonons at large couplings or small phonon frequencies.
To overcome this restriction we will replace the standard Fock basis with a more general construction.
For a single harmonic oscillator (everything generalizes immediately to multiple oscillators at different lattice sites) the Fock states with fixed boson number are the states $|n\rangle = (b^\dagger)^n/\sqrt{n!} |\mathrm{vac}\rangle$.
If a force, say $-\sqrt{\epsilon_p \omega_0} (b^\dagger + b)$, is applied the average number of bosons in the groundstate is given by $g^2=\epsilon_p/\omega_0$. The larger $\epsilon_p$ or the smaller $\omega_0$, the more bosons are needed to account for the oscillator elongation $\langle b^\dagger + b\rangle =2g$.
On the other hand, we know that the groundstate is simply a coherent state $|g\rangle_c = \hat{S}(g) |\mathrm{vac}\rangle$, obtained from the vacuum with the shift operator $\hat{S}(\alpha) = \exp(\alpha b^\dagger - \alpha^* b)$.

In the polaron problem the effective force exerted on the oscillator, hence its state, changes as the electron moves. 
We therefore should try to find states covering the entire range of oscillator elongations.
This motivates the following construction of shifted oscillator states (SOS):
To a given parameter $\sigma>0$, 
start with the coherent states $|n\sigma\rangle_c = \hat{S}(n \sigma) |\mathrm{vac}\rangle$. Orthogonalize these states with the Gram-Schmidt procedure, in the given order, to obtain the SOS $|n\rangle_\sigma$. The state $|0\rangle_\sigma=|0\rangle_c$ remains the vacuum state, while the higher states are a complicated mixture of coherent states. 
The Fock states are recovered in the limit $\sigma \to 0$, when $\hat{S}(\sigma) \simeq 1 + \sigma(b^\dagger-b)$.
More properties of the SOS are given in the appendix (Sec.~\ref{sec:app}).

The average elongation and energy of the SOS $|n\rangle_\sigma$ grows with $n$, so that they can replace the Fock states in the VED. 
The optimal parameter $\sigma$ can be determined variationally through minimization of the groundstate energy.
For our purposes, the choice $\sigma=g/N_\mathrm{gen}$ is close to optimal,
and was used in all calculations.
For larger coupling $g$, hence larger optimal $\sigma$, the SOS account for larger oscillator elongations.
Conversely, if the maximal number of states $N_\mathrm{gen}$ per oscillator is increased and $\sigma$ decreases, the SOS sample ever finer details of the oscillator wavefunctions.
In fact, since in our calculations always $\sigma \lesssim 1$,
the states $|n\sigma\rangle_c$ overlap in position space, 
and the SOS do not resemble coherent states after orthogonalization.
Instead they interpolate between `quantum' Fock states generated by $b^\dagger$ and `classical' coherent states generated by $\hat{S}(\alpha)$.
Note that a construction with a fixed small number of coherent states
would have to fail: For small $\omega_0$ it is not possible to represent all oscillator wavefunctions -- with small width and different positions -- if only few states are available. For the SOS the necessary increase in accuracy is accomplished by construction if $N_\mathrm{gen}$ grows.

For the polaron problem, the gain in efficiency obtained using SOS is demonstrated in Tab.~\ref{fig:accuracy}.
Only this efficiency gain allows us to perform calculations for frequencies $\omega_0 \ll 0.1$ at $\lambda = 1$,
extending the accessible parameter range by more than one order of magnitude into the adiabatic regime.
The present calculations, down to $\omega_0=0.01$, require $N_\mathrm{gen} \le 23$, and could all be performed on small desktop computers.
The validity of our implementation was checked against calculations from Diagrammatic Quantum Monte Carlo (data for the 3D polaron mass at $\omega_0=0.5$ were provided by A.~S.~Mishchenko).

Coherent states have been used in variational~\cite{FILTB94,RBL98,Ba07} and numerical approaches to the Holstein model~\cite{CDMP04,DCMP05}.
The latter two references describe a construction similar to ours.
It appears that our construction is simpler, in that it avoids variational optimization of coherent state parameters, while the additional virtues of the VED construction allow us to perform calculations for 1D and 3D systems in the far adiabatic regime.

\begin{table}
  \begin{tabular}{cccc}
    $N_\text{gen}$  & $N$ & $E$ (SOS) & $E$(Fock) \\\hline
    11   &     5063 & -2.53737656 & -2.4463726 \\
    12   &    10391 & -2.53776224 & -2.4588902 \\
    13   &    21247 & -2.53792335 & -2.4702848 \\
    14   &    43310 & -2.53798172 & -2.4807877 \\
    15   &    88052 & -2.53800006 & -2.4904380 \\
    16   &   178617 & -2.53800511 & -2.4991678 \\
    17   &   361644 & -2.53800635 & -2.5068954 \\
    18   &   731027 & -2.53800663 & -2.5135778 \\
    19   &  1475635 & -2.53800668 & -2.5192232 \\
    20   &  2975103 & -2.53800669 & -2.5238825
  \end{tabular} 
\caption{Variation of numerical groundstate energy $E$ for $\omega_0=0.1$, $\lambda=1$ in 1D, with the number of iterations $N_\text{gen}$ in the VED state selection, and corresponding number $N$ of phonon states per electron site in the Hilbert space. With Fock states, convergence is not achieved. With SOS, $6$ digits of the energy are converged already for $N=10^5$ states.
The groundstate energy on the infinite lattice is obtained with a relative error of less than $10^{-8}$.}
\label{fig:accuracy}
\end{table}

\section{Polaron mass}
\label{sec:mass}

The fundamental quantity characterizing polaron formation is the polaron mass $m$ defined by
\begin{equation}\label{mass}
 m^{-1} = \frac{d^2}{d|\mathbf{k}|^2} E(\mathbf{k}) \Big|_{\mathbf{k}=0} \;,
\end{equation}
where $E(\mathbf{k})$, the polaron dispersion, is the groundstate energy with momentum $\mathbf{k}$.
For a free electron with $\epsilon_p=0$, the mass is $m_0=1/(2t)$.
With increasing coupling, the polaron mass $m$ grows, and the mass renormalization $M = m/m_0$ becomes larger than unity.
Our numerical method works on the infinite lattice with continuous momentum $k$, which allows for a direct evaluation of the $k$-derivative in Eq.~\eqref{mass}.

In the antiadiabatic strong coupling limit ($\omega_0 \to \infty$ with $g^2$ fixed) the polaron mass is given by the asymptotic Lang-Firsov formula $M \simeq \exp(g^2)$,
independent of dimension.
In contrast, the behavior in the adiabatic regime ($\omega_0 \ll t$) depends on dimension, since it is governed by the static limit $\omega_0=0$.
In the static limit the polaron in 1D has infinite mass for all $\lambda>0$, while a phase transition -- the self-trapping transition -- from the free electron ($M=1$) to a polaron ($M=\infty$) occurs at a finite coupling strength $\lambda_c \approx 0.9$ in 3D.
Since this difference is the most significant aspect of polaron formation in different dimensions,
we leave out the 2D case here and concentrate on the 1D and 3D situation.

\begin{figure}
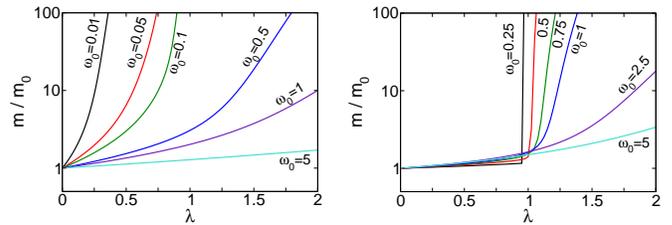

\includegraphics[width=0.48\linewidth]{Fig1a.eps}\hfill
\includegraphics[width=0.48\linewidth]{Fig1b.eps}
\caption{(Color online) Polaron mass as a function of coupling for different phonon frequency, in 1D (left panel) and 3D (right panel), for $\omega_0$ as indicated.}
\label{fig:mvsl}
\end{figure}

In Fig.~\ref{fig:mvsl} we show how the polaron mass at finite $\omega_0$ interpolates between the two limits.
Both for 1D and 3D the polaron mass shows the asymptotic Lang-Firsov behavior for $\omega_0 \gg t$.
In 1D, the curves get steeper for smaller $\omega_0$, while keeping their functional form.
The curves for $\omega_0 \ll t$ and $\omega_0 \gg t$ differ only quantitatively.
In particular, a singular value of $\lambda$ (or $\omega_0$) associated with a qualitative change does not exist.
In this sense the 1D behavior is scale-free.

Quite different in 3D: The curves develop a sharp bend close to $\lambda=1$ if $\omega \lesssim 0.5$, whose position converges to $\lambda_c$ for even smaller $\omega_0$.
In this case a small change of $\lambda$, in the vicinity of $\lambda_c$, results in a dramatic change of $m$.
Although a true phase transition does not occur~\cite{GL91} for any $\omega_0 > 0$,
already at $\omega_0 = 0.5$ the behavior of the polaron mass is, for all practical purposes, indistinguishable from that for a phase transition.
In the adiabatic 3D situation one deals either with a quasi-free electron or a very heavy polaron.

\begin{figure}
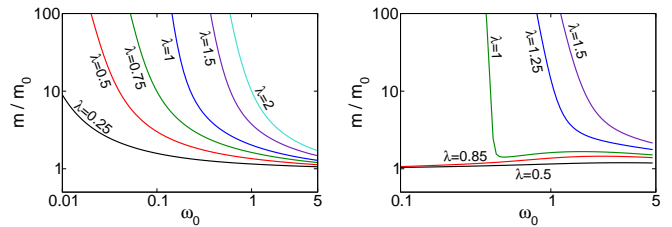

\includegraphics[width=0.48\linewidth]{Fig2a.eps}\hfill
\includegraphics[width=0.48\linewidth]{Fig2b.eps}
\caption{(Color online) Polaron mass as a function of phonon frequency for different coupling, in 1D (left panel) and 3D (right panel),
for $\lambda$ as indicated.}
\label{fig:mvso}
\end{figure}

Plotting the polaron mass as a function of phonon frequency (Fig.~\ref{fig:mvso}) demonstrates the behavior discussed here in a very clear way.
In 1D the mass diverges for all $\lambda>0$ if $\omega_0$ goes to zero.
In 3D the mass diverges only if $\lambda>\lambda_c$, while it approaches $m_0$ if $\lambda<\lambda_c$.
Just as before, now a small change in $\omega_0$ results in a large change of $m$ for $\lambda$ close to $\lambda_c$.
In both situations the mass at finite frequency converges correctly to that in the static limit.

\section{Polaron radius}\label{sec:radius}

The different behavior of polaron formation in 1D and 3D,
characterized through the polaron mass in the previous section,
can also be studied for the size of a polaron, given by the extension of the phonon cloud surrounding the electron.
Using the electron-phonon correlation function
$ \chi(\mathbf{r}) = \sum_\mathbf{n} \langle \psi_0|
(b^\dagger_{\mathbf{n}+\mathbf{r}}+b_{\mathbf{n}+\mathbf{r}})
c^\dagger_\mathbf{n} c_\mathbf{n} |\psi_0\rangle$,
where $|\psi_0\rangle$ is the groundstate at momentum $\mathbf{k}=0$, 
the polaron radius is defined as
\begin{equation}\label{polrad}
  R = \left[ \frac{ \sum_\mathbf{r} |\mathbf{r}|^2 \chi(\mathbf{r})
    }{2 D \sum_\mathbf{r} \chi(\mathbf{r}) }   \right]^{1/2} \;.
\end{equation}
Note that for the Holstein model, $\sum_\mathbf{r} \chi(\mathbf{r}) = 2\sqrt{\varepsilon_p/\omega_0}$.

\begin{figure}
\includegraphics[width=0.9\linewidth]{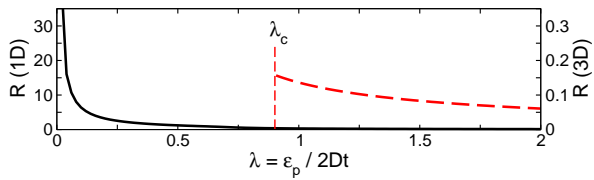}
\caption{(Color online) Polaron radius as a function of coupling in the static limit, in 1D and 3D. Note the different scaling of the $R$-axis on the left and right. The vertical dashed line indicates the critical coupling $\lambda_c$ for the 3D adiabatic self-trapping transition.}
\label{fig:static}
\end{figure}

For the static limit, the polaron radius is shown in Fig.~\ref{fig:static}.
We observe that the radius changes continuously in 1D.
In 3D, the radius is finite only above the self-trapping transition at $\lambda_c \approx 0.9$, but the self-trapped polaron is a very small particle, with $R \lesssim 0.2$.
Note that in the static limit only the polaron radius provides information about the polaron properties, as the mass is infinite above the self-trapping transition and does not change.

\begin{figure}
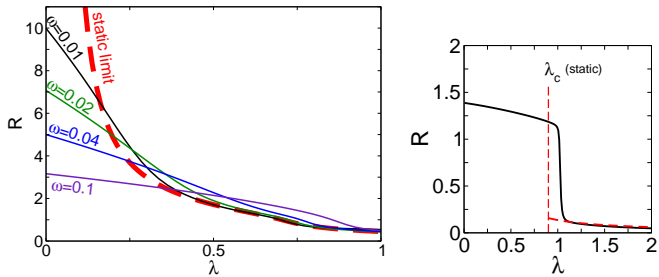

\includegraphics[width=0.58\linewidth]{Fig4a.eps}\hfill
\includegraphics[width=0.38\linewidth]{Fig4b.eps}
\caption{(Color online) Polaron radius as a function of coupling.
Left panel: For different phonon frequencies in the adiabatic regime in 1D.
Right panel: In 3D for $\omega_0=0.5$.
The red dashed curves give the corresponding polaron radius in the static limit (from Fig.~\ref{fig:static}),
the dashed vertical line gives $\lambda_c$ for the static limit.
}
\label{fig:rad}
\end{figure}

Turning to finite phonon frequencies, 
one should first understand that the polaron radius remains finite in the weak coupling limit.
The reason is that any additional phonon requires a finite energy $\omega_0$,
which suppresses phonon excitations far away from the electron where
the energy increase is not compensated by electron-phonon interaction.
The limiting value of $R$, as $\lambda$ approaches zero, is given by~\cite{RBL99}
\begin{equation}
  \lim_{\lambda \to 0} R = \sqrt{\frac{t}{\omega_0}} \;.
\end{equation}
This expression implies that a large polaron can only exist for sufficiently small $\omega_0$, e.g. $R>5$ requires that $\omega_0 < 0.04$ independent of coupling strength. 
A different question is the existence of a large and heavy polaron.
As the results for the static limit indicate, such a polaron can only exist in 1D, at very small $\omega_0 \ll t$ and small coupling $\lambda \lesssim 0.5$.

In Fig.~\ref{fig:rad} we show the polaron radius in 1D and 3D.
One should note that the evaluation of Eq.~\eqref{polrad} is not possible with extremely high accuracy, due to the summation in the numerator containing $|\mathbf{r}|^2$.
In fact, the expression is not even well conditioned since states with arbitrarily small energy difference can have totally different $R$ (just excite a phonon very far from the electron).
For the present purpose, we calculate $R$ under the assumption that $\chi(\mathbf{r})$ decays exponentially for sufficiently large $|\mathbf{r}|$, and the resulting error is about $3\%$ in the most difficult situation ($\omega_0=0.01$, $\lambda \simeq 0.5$).

As expected, $R \ll 1$ in 3D if $\lambda \gtrsim \lambda_c$.
In 1D we find that, at the smallest frequency $\omega_0 = 0.01$ studied,
a polaron with radius $R>2$ ($R>3$) exists for $\lambda < 0.43$ ($\lambda < 0.3$).
If we compare these values with the polaron mass in Fig.~\ref{fig:mvsl},
we find, e.g. for $\lambda = 0.25$, a large and heavy polaron with $R>4$, $M>10$.

\begin{figure}
\includegraphics[width=0.9\linewidth]{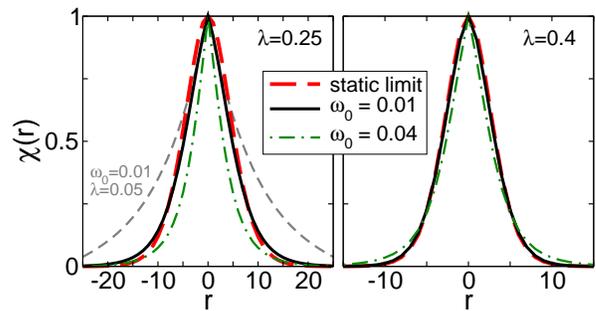}
\caption{(Color online) Electron-phonon correlation function $\chi(r)$ in 1D for $\omega_0=0.01, 0.04$,
in comparison to the static limit (red curve).
The gray dashed curve in the left panel shows $\chi$ for $\omega_0=0.01$, at weaker coupling $\lambda=0.05$.}
\label{fig:1DCorr}
\end{figure}

To support our claim that polarons for these parameters 
correspond to what is often called an adiabatic polaron in the literature
we show in Fig.~\ref{fig:1DCorr} the electron-phonon correlation function $\chi(r)$ in 1D for $\omega_0=0.01$, $\lambda=0.25$ in comparison to that for the static limit, which closely resemble each other.
The agreement becomes even better for $\lambda=0.4$,
although then the polaron with $R\gtrsim 2$ is not really large.
Note that in order to find such agreement one has to use phonon frequencies as small as $\omega_0=0.01$. Already at $\omega_0=0.04$ the curves do not fit.

\section{Survey of the 1D polaron}
\label{sec:1d}

\begin{figure}
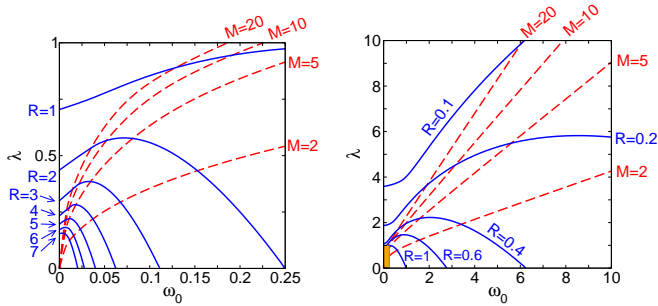

\includegraphics[width=0.48\linewidth]{Fig6a.eps}\hfill
\includegraphics[width=0.48\linewidth]{Fig6b.eps}
\caption{(Color online) Curves of constant mass, and curves of constant radius in 1D. The left panel is a magnification of the adiabatic regime $\omega_0\le 0.25$, which corresponds to the small orange rectangle in the right panel.}
\label{fig:1Dchart}
\end{figure}

\begin{figure}
\centering
\includegraphics[width=0.75\linewidth]{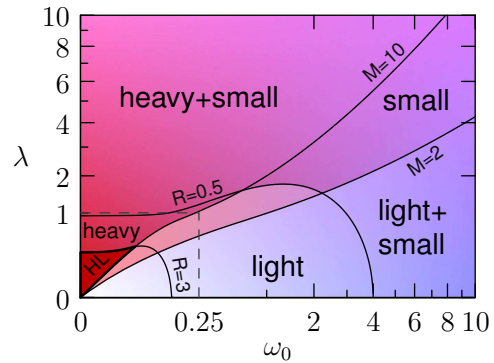}
\caption{(Color online) Overview of different polaron regimes in 1D.
Only for the notation in this figure, we call a polaron light (heavy) if $M<2$ ($M>10$),
and large (small) if $R>3$ ($R<0.5$).
With this terminology, heavy large polarons are found in the region ``\textsf{HL}''.
The gray dashed rectangle gives the parameter region shown in the left panel in Fig.~\ref{fig:1Dchart}. 
For better visibility, the region of smaller $\omega_0$, $\lambda$ values is magnified through a
continuous variation of the scaling along each axis.
}
\label{fig:1DArt}
\end{figure}

We have seen in the previous two sections that in the adiabatic regime in 3D 
light polarons evolve into very heavy and very small polarons through a rapid crossover.
Only in 1D a continuous renormalization of the polaron mass and radius occurs in an extended region of parameter space.
A map of that region is given in Fig.~\ref{fig:1Dchart}
where we plot curves of constant mass and of constant radius.
If we assign names to certain regimes bordered by the $M=2, 10$ and $R=0.5, 3$ curves,
we get Fig.~\ref{fig:1DArt}, which indicates the tendencies in the evolution from a light to a heavy, or a large to a small polaron with changing model parameters. 
Independent of the precise choice of wording, large and heavy polarons exist for sufficiently small $\omega_0$. 
The existence of such polarons is specific for 1D,
and a similar regime does not appear in 3D, where heavy polarons are always small as we discussed in the previous section.
Even in 1D, it requires very small phonon frequencies.
In combination with the comparison of the electron-phonon correlation function in Fig.~\ref{fig:1DCorr} we can therefore refine the conventional adiabaticity condition $\omega_0 < t$ to a much stricter statement, namely $\omega_0/t \ll 0.1$ or $\omega_0/t \lesssim 0.01$.
The specific 1D physics of large heavy polarons thus occurs only in a tiny region of parameter space.
Turning to larger phonon frequencies, the polaron is always small.
For $\omega_0/t \gg 1$, the polaron is almost point-like.
Rearranging Eq.~\eqref{polrad} gives $\chi(0)/\sum_\mathbf{n} \chi(\mathbf{n}) \ge 1-R^2$,
so that e.g. for $R=0.5$ at least $75\%$ of the total lattice displacement are located at the electron. 
In this limit, the only significant polaron parameter is the average number of phonons $g^2=\epsilon_p/\omega_0$ -- instead of the two parameters $\omega_0$ and $\lambda$ --, the mass is given by the asymptotic Lang-Firsov expression $M = \exp(g^2)$, and the differences between dimensions vanish.

\section{Optical response in 1D}\label{sec:opt}
The physics of adiabatic, or large and heavy, polarons is essentially different from that of small polarons in the antiadiabatic strong coupling regime. 
The difference extends beyond the differences in mass or size of the polaron discussed so far,
but affects its very structure, 
and thus relates to the question in which sense a polaron can be understood as an itinerant quasiparticle~\cite{BEMB92,MR97,MR97err}. For a characterization of the quasiparticle properties we can study its response to electric fields.
The (regular part of the) optical conductivity $\sigma_\mathrm{reg}(\omega)$ is defined through 
\begin{equation}\label{sigmareg}
  \sigma_\mathrm{reg}(\omega) = 
  \pi \sum_{l>0} \frac{\langle \psi_l | \hat{\jmath} |\psi_0\rangle |^2}{\omega_l}
  \,[\delta(\omega - \omega_l) + \delta(\omega+\omega_l)] \;,
\end{equation}
with the current operator $\hat{\jmath} =  \mathrm{i} t \sum_n \,
  c^\dagger_{n+1} c^{}_n - c^\dagger_n c^{}_{n+1}$
  and the eigenstates $|\psi_l\rangle$ with excitation energy $\omega_l$.
Further information is contained in the f-sum rule
\begin{equation}\label{fsumrule}
  2\pi \mathcal{D} + 2 \int_0^\infty \! \sigma_\mathrm{reg}(\omega) d\omega =  
  - \pi \, E_\mathrm{kin} \;,
\end{equation}
where
$E_\mathrm{kin} = -t \langle \psi_0|\sum_n c^\dagger_{n+1} c^{}_n +
c^\dagger_n c^{}_{n+1} |\psi_0 \rangle$ is the kinetic energy.
As a basic transport quantity here the Drude weight $\mathcal{D}$ appears,
which is related to the polaron mass $m$ by $\mathcal{D}=1/(2m)$.

The reader should be aware that in a numerical calculation of  $\sigma_\mathrm{reg}(\omega)$ it is impossible to capture the exact position and weight of all excited states.
For our results, the resulting error in integrated quantities, e.g. the first few moments of $\sigma_\mathrm{reg}(\omega)$, is negligible. 
For $\sigma_\mathrm{reg}(\omega)$ itself 
a compromise between finite spectral resolution and the finite level spacing in the numerical Hilbert space has to be found. Consequently, the shape, envelope and peak positions in the pictures for $\sigma_\mathrm{reg}(\omega)$ (see Fig.~\ref{fig:optics} below) are correct, but it should be clear that tiny spikes on top of the curves are a kind of numerical noise. 

\begin{figure}
\includegraphics[width=0.9\linewidth]{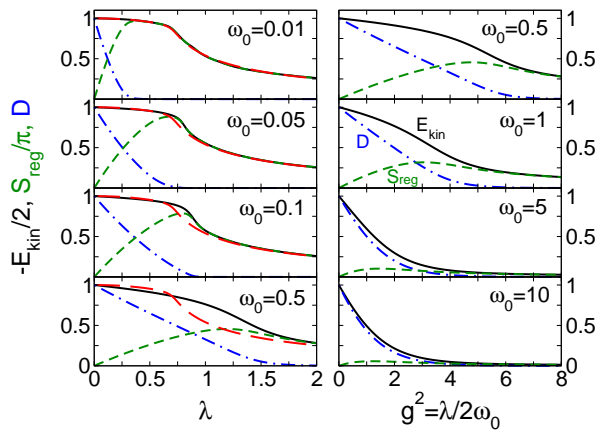}	
\caption{(Color online) Depiction of the f-sum rule, with kinetic energy $E_\mathrm{kin}$ (solid), integrated regular conductivity $S_\mathrm{reg}=\int_0^\infty \sigma_\mathrm{reg}(\omega') d\omega'$ (dashed), and Drude weight $\mathcal{D}$ (dash-dotted) in 1D.
The long-dashed red curve in the left column gives the kinetic energy in the static limit for comparison.
For $\omega_0=0.01$, it lies on top of the $E_\mathrm{kin}$-curve.
}
\label{fig:sumrule}
\end{figure}

We show the evaluation of the f-sum rule in Fig.~\ref{fig:sumrule}.
Note that, working on an infinite lattice, the f-sum rule is satisfied in all parameter regimes.
In particular, the Drude weight obtained either from the f-sum rule via $\sigma_\mathrm{reg}(\omega)$ or from the polaron mass differs by less than $10^{-4}$.

As we have discussed above,
the polaron in the antiadiabatic strong coupling regime with $R<1$ attains a large mass renormalization only through local phonon excitations tightly bound to the electron. Such a polaron is always a very small
particle, with strong correlations between electron and phonon
dynamics. We therefore expect that, with increasing coupling, the kinetic energy and Drude weight simultaneously become small as the polaron gets less mobile,
 and that at the same time the optical conductivity decreases since a very small object is less susceptible to optical fields.
As this expectation is confirmed in Fig.~\ref{fig:sumrule} (right bottom panel),
we can say that in this type of polaron the electron is dressed with phonons, resulting in a
new polaronic quasiparticle emerging as a joint
electron-phonon entity.

The physics is essentially different in the adiabatic regime of heavy but large polarons.
For such adiabatic polarons electron and phonon dynamics
partially decouple, anticipating the static limit where the electron
moves in the rigid phonon configuration of the lattice displacement.
That there translational symmetry is broken
and the polaron has infinite mass
is equivalent to the fact that states of a static oscillator with
different position, and vanishing width, have zero overlap.
At small but finite $\omega_0$ the translated phonon configurations
have small overlap, and the polaron acquires a finite but eventually large
mass even for large radius.
That the radius remains large means that in contrast to the
antiadiabatic situation the mass increase is not so much a result of
dressing the electron with phonons, but rather of the immobility of
the extended lattice displacement as a whole.
The electron motion relative to the lattice displacement is less
suppressed, which results in the large radius.
Again, we see this behavior in the f-sum rule in Fig.~\ref{fig:sumrule} (left top panel).
Now, even if the Drude weight close to zero indicates the immobility of the polaron, the kinetic energy and optical conductivity remain large.
The opposite behavior of these quantities can be attributed to the different constituents of the polaron.
The Drude weight is small since the phonon part suppresses motion of the 
whole polaron,
but the kinetic energy is large due to the electron motion relative to the phonons. Transitions between different electronic states in the almost rigid phonon background give a large contribution to the optical conductivity.
If electron and phonon motion decouple in this way an adiabatic polaron can be understood as a quasiparticle only with certain restrictions.
The polaron is rather a composite electron-phonon object, where electron and phonons influence each other but remain separable, instead of that joint electron-phonon entity found in the antiadiabatic regime where the electron and the phonons lose their independence in favor of a new quasiparticle.
As the f-sum rule shows this difference persists through the entire coupling range.

\begin{figure}
\includegraphics[width=0.9\linewidth]{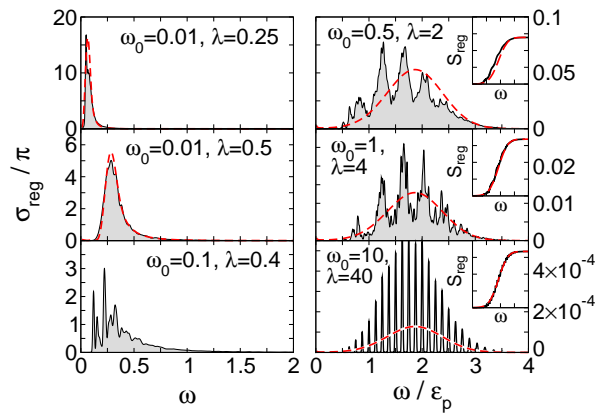}	
\caption{(Color online) Optical conductivity $\sigma_\mathrm{reg}(\omega)$ in 1D.
Left column: For small $\omega_0$. 
The red dashed curves in the upper two panels give the conductivity in the static limit $\omega_0=0$, with same $\lambda$.
Right column: For large $\omega_0$.
The red dashed curves show the approximate Gaussian expression for the conductivity obtained in small-polaron theory~\cite{RH67,Emi93}.
The insets compare the integrated conductivity $S_\mathrm{reg}(\omega)= \int_0^\omega \sigma_\mathrm{reg}(\omega') d\omega'$ (in arbitrary units).
}
\label{fig:optics}
\end{figure}

Differences between the adiabatic and antiadiabatic regime appear also in the optical conductivity (see Fig.~\ref{fig:optics}).
At large phonon frequencies and strong coupling the optical conductivity is dominated by phonon-assisted transitions to excited polaron states, which change the average phonon number by an integer.
The single peaks are thus separated by $\omega_0$, and resolve the Poisson distribution of phonons in the polaron groundstate.
The skewed Gaussian envelope of $\sigma_\mathrm{reg}(\omega)$ is a signature of antiadiabatic small polarons~\cite{RH67,Emi93}.
Note that the integrated optical conductivity $S_\mathrm{reg}$ is small in this regime.
For intermediate phonon frequencies complicated structures develop,
with a contribution from different polaron bands whose width is no longer small compared to $\omega_0$.
In the far adiabatic regime the basic mechanism for optical absorption has changed. It is now dominated by electronic transitions in the potential well generated by the lattice displacement.
Of course, a phonon admixture is present for any finite $\omega_0$.
Only in the static limit the phonon configuration remains unchanged during an electronic transition, and the Franck-Condon effect is fully realized.

To summarize, the change in the optical conductivity with phonon frequency reflects very clearly the two different time scales realized in the polaron problem. Depending on whether the electron or the phonons set the fastest time scale, we observe different physics.
In the antiadiabatic regime we find a well-defined polaronic quasiparticle whose behavior is governed by local phonon excitations dressing the electron and the arising strong correlations between electron and phonon dynamics.
Or we find, in the adiabatic regime, an electron moving in a phononic background whose properties depend only on the average electron motion, while stronger correlations are missing.
These differences and their consequences for polaron transport and dynamics have been studied here in terms of the optical conductivity and f-sum rule.

\section{Conclusion and Outlook}
\label{sec:conclusion}

A description of lattice polaron formation 
requires the treatment of the two different time-scales of electron and phonon dynamics, their mutual influence and accompanying renormalization, and the non-trivial quantum correlations that develop.
The adiabatic regime is especially difficult, which is typical for many problems
where one tries to perform the classical limit for some part of a system while keeping the full quantum dynamics of the remainder.

For the Holstein model one is in the fortunate position that physical concepts or approximate theories can be checked against very accurate data provided by advanced numerical calculations. 
In that way the validity of important ideas, potentially with generalizations to less tractable situations, can be assessed, and some of the controversies in the polaron literature may be resolved.
In the present contribution we develop our physical discussion along the lines indicated by such accurate data. The new quality of this contribution is our ability to cover the entire coupling and frequency range, in particular the far adiabatic regime including heavy large polarons in 1D.
This progress becomes possible by the new numerical method introduced here.

Our results support a particular picture of polaron formation, which is in agreement only with parts of the literature. Let us summarize some important findings.
(I) Although all groundstate quantities vary continuously with coupling and phonon frequency, 
the steady change in 1D is replaced with a rapid crossover in the adiabatic regime in 3D, which closely resembles a true phase transition.
While no characteristic energy scale exists in 1D,
it is set in 3D by the critical coupling $\lambda_c$ of the self-trapping transition in the static limit. \\
(II) All quantities converge (or diverge) to the static limit in a simple and clear way.
Nevertheless quantitative agreement is obtained only for very small phonon frequencies $\omega_0/t \ll 0.1$.\\
(III) Only for such small phonon frequencies heavy and large polarons occur in 1D (e.g. $\omega_0 < 0.02$ for $M>5$ and $R>4$). A large radius requires a small phonon frequency in any case.\\
(IV) Small and large polarons have significantly different transport and optical properties.
A large and heavy polaron is characterized through a small Drude weight but large optical absorption.
A small and heavy polaron has small Drude weight and small optical absorption. \\
(V) The small-polaron approximation for the optical conductivity is strictly valid only for very large phonon frequencies, while in general pronounced structures distort the Gaussian envelope.
The overall shape of the envelope however remains comparable also for smaller phonon frequencies.

These findings suggest certain physical interpretations.  
First, with all necessary caution, it may make sense to think of a self-trapping transition in 3D even at finite phonon frequency. There is nothing comparable in 1D.
Second, the notion of an itinerant polaronic quasiparticle is to be used only with care:
The nature of the quasiparticle differs at large and small phonon frequency (cf. Sec.~\ref{sec:opt}).
Third, Holstein polarons tend to be small objects. 
This should largely prevent polaron dissociation, which may occur in a kind of polaronic Mott transition if the phonon clouds overlap at finite polaron density.
It also prevents instabilities in anisotropic systems with a preferred direction of motion, where only the large adiabatic quasi-1D polarons are destabilized by electron hopping perpendicular to this direction~\cite{Emi86,AFT08}.

In principle, our results allow for direct comparison with, and hence validation of, approximate polaron theories. We have not tried to do so here, since we want our results to tell their own tale.
Our discussion should have shown that the numerical data itself do not prevent a physical interpretation.
But the interpretation adheres to direct terms -- the polaron is heavy or light, small or large --, and simple pictures --  such as a phonon-dressed electron vs. an electron moving in the potential generated by the lattice displacement. There is no necessity to invoke complicated constructions which do not find their counterpart in the few quantities truly characterizing the polaron.
 
The present problem, i.e. the single Holstein polaron at zero temperature,
is treated here exhaustively. Of course, it is possible to go to even smaller phonon frequencies, larger couplings or other more extreme situations, but new physical behavior will not occur.
There are however many other topics in polaron physics that remain unsettled.
With a few singular exceptions none of them have been accessed with the same rigor as the present problem. 
Within the framework of the Holstein model, transport at finite temperature and collective effects at finite density are probably the two most important issues.
Proceeding beyond the Holstein model, in particular on trying to extend the discussion to polarons in novel materials such as graphene~\cite{CB08}, new questions emerge.
These include polaron formation in the presence of impurities~\cite{MNAFDCS09} and disorder~\cite{GJ80,BF}, the influence of phonon dispersion, of different types of electron-phonon coupling, of lattice geometry, anisotropy~\cite{AFT08} or confinement,
or the dynamics of polaron formation~\cite{KT07} out of equilibrium, realized e.g. in quantum dots or molecular aggregates. 
Some of these questions are directly connected with the present study.
Does, e.g., longer-ranged electron-phonon interaction favor the formation of large and heavy polarons? 
Although the polaron becomes larger then, its mass will decrease, and the existence of a large heavy polaron with corresponding optical signatures may crucially depend on a number of material parameters. 
The classification of these parameters and of their influence on polaron properties is central to giving studies such as the present one relevance beyond the rigorous treatment of a model.
To make such investigations possible is certainly the motivation to develop powerful high-precision numerics for the polaron problem, while their application provides us with the necessary information to understand real polarons and the models to describe them.


\section*{Acknowledgments}
This work was supported by the US Department of Energy, Center for Integrated Nanotechnologies, at Los Alamos National Laboratory (Contract DE-AC52-06NA25396) and Sandia National Laboratories (Contract DE-AC04-94AL85000).
A.A. and H.F. are grateful for hospitality provided at the Los Alamos National Laboratory.

\appendix

\section{Shifted Oscillator States}\label{sec:app}

The SOS $|n\rangle_\sigma$, with $n\ge 0$, form a Hilbert space basis for a single harmonic oscillator.
For $\sigma \ne 0$, they are characterized through two properties: (i) orthonormality $\langle n|m\rangle_\sigma = \delta_{nm}$,
(ii) the $n$-th SOS is a linear combination 
\begin{equation}
 |n\rangle_\sigma = \sum_{m=0}^n c_m^{(n)} |m\sigma\rangle_c
\end{equation}
of the $n+1$ coherent states $|m\sigma\rangle_c$, $0 \le m \le n$.
In particular, $|0\rangle_\sigma = |0\rangle_c = |\mathrm{vac}\rangle$.

These conditions imply that the SOS can be constructed through Gram-Schmidt-orthogonalization of the equally spaced coherent states $|m\sigma\rangle_c$, in the order of increasing $m=0,1,2,\dots, \infty$.
Moreover, the operator $b$ is given by a upper triangular matrix in the SOS basis.
It is possible to obtain explicit expressions for the coefficients 
\begin{equation}
c_m^{(n)}= \frac{(-\alpha)^{n-m}}{\Big[ \prod\limits_{k=1}^n (1-\alpha^{2k}) \Big]^{1/2}} \, \genfrac{[}{]}{0pt}{}{n}{m}_{\alpha^2} \;,
\end{equation}
where $\alpha =e^{-|\sigma|^2}$ and the q-binomials $\left[\begin{smallmatrix} n \\ m \end{smallmatrix} \right]_q$ (for a definition, see e.g. Ref.~\onlinecite{hyper}) are polynomials in $q$.
All relevant matrix elements can also be computed with a numerical Gram-Schmidt procedure.

The above definition applies to all $\sigma \ne 0$.
The limit $\sigma \to 0$ exists, and the SOS reproduce the Fock states in this limit. In the opposite limit $\sigma \to \infty$,
they become asymptotically equal to the coherent states $|m\sigma\rangle_c$,
since the overlap between different coherent states is small.

From the explicit expressions for $c_m^{(n)}$ we find that $\langle n|b|n\rangle_\sigma = n \sigma$.
For real $\sigma$, it follows that the oscillator elongation in the $n$-th SOS is given by $2 n \sigma$.
Thus, SOS for $\sigma>0$ ($\sigma<0$) should be used for states with positive (negative) elongation.

\begin{figure}
\includegraphics[width=0.9\linewidth]{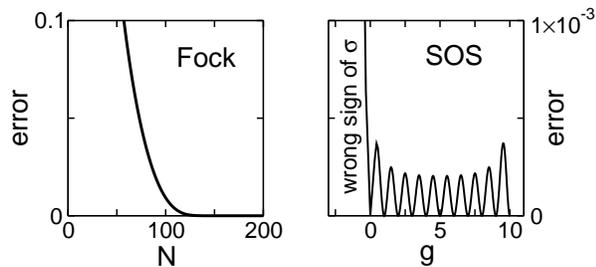}	
\caption{Absolute error of the groundstate energy of the Hamiltonian~\eqref{HamOsc} with $\omega_0=0.01$, for a maximal $g^2=100$, as explained in the text.
Left panel: Error using $N+1$ Fock states.
Right panel: Error using $N=10$ SOS with $\sigma=g^\mathrm{max}/N=1$, as a function of $g$.
}
\label{fig:SOS}
\end{figure}

To quantify the benefit of SOS over Fock states for groundstate calculations, we consider a shifted oscillator with Hamiltonian
\begin{equation}\label{HamOsc}
H_\mathrm{osc}=\omega_0 b^\dagger b^{} - g \omega_0 (b^\dagger+b^{}) \;.
\end{equation}
The groundstate of this Hamiltonian, with energy $-\epsilon_p=-g^2 \omega_0$, is a coherent state $|g\rangle_c$, which contains $g^2$ bosons on average.
In a numerical calculation an approximate groundstate of $H_\mathrm{osc}$ is calculated in a truncated Hilbert space, e.g. consisting of the first $N+1$ Fock states. The corresponding error in the groundstate energy is shown in Fig.~\ref{fig:SOS}.
As expected, the error is large when $N < g^2$.

Using SOS instead of Fock states, the error depends on $\sigma$.
In the present example, the error is zero for $\sigma=g/N$, 
since the coherent state $|g\rangle_c$ is a linear combination of the first two SOS.
For a meaningful error estimate we therefore fix 
$\sigma = g^\mathrm{max}/N$ for the maximal $g^\mathrm{max}$ considered.
The virtue of the SOS is that the error is small also for all intermediate $0 < g < g^\mathrm{max}$ (see Fig.~\ref{fig:SOS}).
Note that the error drops to zero at those values of $g$ for which the true groundstate is contained in the truncated Hilbert space spanned by the $N+1$ SOS. 

We also see how the SOS prefer positive elongations for $\sigma>0$,
and the error becomes large if $\sigma$ and $g$ have opposite sign.
In a real calculation, the variational determination of the parameter $\sigma$ would prevent this situation.
For the Holstein model~\eqref{Ham}, only positive elongations occurs and always $\sigma>0$.
Note that in the worst case the optimal $\sigma$ is zero, and 
the Fock states are recovered. Therefore, the introduction of SOS can only improve groundstate results obtained with Fock states.
In the given example, the error with $10$ SOS is one order of magnitude smaller than that with $100$ Fock states.

The SOS construction can be modified in various ways.
For example, if oscillator states with large positive and negative elongations should be represented, one can apply a global shift $\hat{S}(\xi)$,
constructing the SOS from coherent states $|m\sigma+\xi\rangle_c$.
The additional parameter $\xi$ provides the necessary freedom to deal with such a situation.

We introduced the SOS mainly for oscillator groundstates.
They also work for excited states, but we should distinguish two cases.
First, for a single excited state,
the SOS with variational determination of $\sigma$ will succeed and can provide us with good results even in cases when Fock states fail.
Or, second, we may need all excited states at once, such as in the calculation of the optical conductivity $\sigma_\mathrm{reg}(\omega)$.
In this case, the fundamental problem of any ED calculation is that a finite Hilbert space does not allow
for the representation of infinitely many excited states.
No basis construction can overcome this obstruction.
It is however possible to include sufficiently many states in the Hilbert space,
such that the Hilbert space truncation shows up only through a kind of numerical noise,
while the relevant physical structures are resolved.
For example, the equally spaced peaks in the lower right panel in Fig.~\ref{fig:optics} are physically meaningful, while the tiny spikes in the middle right panel are numerical noise.
The SOS construction guarantees the correct envelope and shape of $\sigma_\mathrm{reg}(\omega)$.
For example, in the middle left panel in Fig.~\ref{fig:optics}, a calculation with Fock states instead of SOS results in a curve that is shifted to smaller energies. Again, the Fock states cannot account for the substantial oscillator elongation in the groundstate and excited states in the adiabatic regime.
To summarize, integrated or averaged properties of excited states are obtained correctly with SOS, but not with Fock states. For correct spectral properties, e.g. peak positions at high resolution, additional effort is needed in any case.

\end{document}